\newcommand{\SD}{$\Sigma_D$ }
\newcommand{\SO}{$\Sigma_O$ }
\newcommand{\bS}{\textbf{S} }

\documentclass[twocolumn,nofootinbib,amsmath,amssymb,floatfix]{revtex4}

\usepackage{graphicx}
\usepackage{bm}

\begin{document}

\title{Diffraction to De-Diffraction}

\author{Vladimir F. Tamari}
\email{vftamari@zak.att.ne.jp}
\thanks{This  paper is presented in its original 1993 form,
except for some minor revisions in some sentences, and was not
presented for publication. The help of D. de Lang in putting this
paper online is gratefully acknowledged.} \affiliation{4-2-8-C26
Komazawa, Setagaya-ku, Tokyo, Japan 154-0012}

\begin{abstract}
De-diffraction (DD), a new procedure to totally cancel diffraction
effects from wave-fields is presented, whereby the full field from
an aperture is utilized and a truncated geometrical field is
obtained, allowing infinitely sharp focusing and non-diverging
beams.  This is done by reversing a diffracted wave-fields'
direction. The method is derived from the wave equation and
demonstrated in the case of Kirchhoff's integral. An elementary
bow-wavelet is described and the DD process is related to quantum
and relativity theories.
\end{abstract}


\date{\today}

\maketitle

\section{Introduction}
The minimization of diffraction effects from wave fields in the
focal plane has been proposed for microwaves by
Schelkunoff~\cite{ref1} and for  light by Toraldo di
Francia~\cite{ref2} but with little practical advantage, since the
non-diffracting portion of the field was very faint, surrounded by
giant side lobes, apart from the enormous practical difficulties
of creating the phase and amplitude changes in the aperture
plane~\cite{ref3}. More recently Durnin~\cite{ref4} has
demonstrated that a $J_0$ field will have a non-diffracting
component, but again the beam, originally emitted by an annular
aperture, is faint and of limited extension, with most of the
radiation dissipated in the side lobes. Tamari~\cite{ref5,ref6}
has qualitatively described a method to totally cancel diffraction
effects (de-diffraction, or DD) from any wave field utilizing the
radiation from the full aperture which can be focused to a sharp
point, or non-diverging beam. In this paper DD will be derived
from first principles in terms of the wave equation and the
concept of the reversal of the field.

While DD fields are of most interest in electromagnetic or
acoustic applications such as in super-resolving imaging or in
non-divergent beams, the method is quite general and can be
adopted to any flow-field whether it is a wave field or not, and
even to electrostatic and gravitational potential fields if the
concept of streamline concentration is used instead of DD.
However, for the purposes or this paper the language of
monochromatic electromagnetic fields will be used, since it is in
such cases that diffraction effects have been studied most
thoroughly in optical and microwave applications.

In section~II, DD fields will be defined and shown to be
theoretically possible solutions to the differential wave
equation. Although it is possible to proceed directly to
de-diffraction theory by field reversal in section~V, the related
concepts of streamlined flow and vorticity will be used in
section~III to describe the transformation of a geometrical field
to a diffracted field, illustrated by Kirchhoff's derivation of
the diffraction integral. In section~IV a `bow wavelet' or dipole
model will be examined, supplanting that of the Fresnel-Huygens
wavelet, the better to describe the field in the immediate
vicinity of the aperture, the crucial area where the diffraction
process is both initiated and from which it could be cancelled.
Some DD experimental results will be described in section~VI and
practical applications of the method discussed in section~VII. The
surprising implications of DD theory as far as both quantum and
general relativity theories are concerned will be discussed in
section~VIII and the paper is summarized in section~IX.

\section{DD: Truncated Geometrical Fields}\label{sec:Sec2}
Let \bS be the propagation (velocity, or Poynting) vectors of
$G(x,y,z,t)$, a general solution to the wave equation
\begin{equation}\label{eq1}
\nabla^2G=\frac{1}{v^2}\frac{\partial^2 G}{\partial t^2}
\end{equation}

In an infinite homogeneous medium free from obstacles or current
sources,

\begin{equation}\label{eq2}
\text{curl \textbf{S}} =0
\end{equation}

\noindent{and} hence the energy streamline $S$ (found by solving
the differential equation $G_y/G_x=\partial y/ \partial x$) and
which are the loci of \bS, will be straight lines. The field  is
said to have normal rectilinear congruence and for our purposes
here such a field can be termed a geometrical~\cite{ref7} field
but \textit{without} the usual restriction that the wavelength
$\lambda= 0$, usually understood by the term. This duplication of
terms has been made on purpose since apart from the wavelength the
fields above obey all the laws of optics and are infinite waves
which fill all of space. Current flows in open water are examples
of such smooth fields. If the field's intensity varies in time,
some examples would be an infinite acoustic or electromagnetic
plane wave or a spherical wave emitted by a point source $P_0$
when the fields are characterized by an even intensity function
across equipotential surfaces (the wavefronts) to which $S$ are
always normal.

\begin{figure}[htb]
\includegraphics[width=8.6cm]{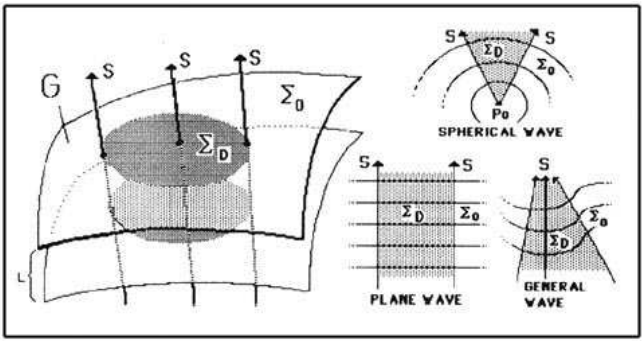}
\caption{\label{fig:1} Various geometrical waves G truncated
within diffraction-free regions \SD, bound by propagation vectors
\bS;  Within \SO the intensity is zero.}
\end{figure}

Referring to Fig.~1, a geometrical wave $G(x,y,z,t)$ is truncated
according to conditions:
\begin{eqnarray}\label{eq3}
G&=&G\;\;\;\; \text{on} \;\; \Sigma_D \nonumber\\
G&=&0\;\;\;\; \text{on} \;\; \Sigma_0
\end{eqnarray}

Where \SD is any limited and  well-defined region of space bounded
by a `wall' of $S$, and \SO is its complementary region, filling
the rest of space outside. Such a wave is, theoretically also a
solution because in \SO  the wave $G(x,y,z,t)=0$ is also a
solution of Eq.~\ref{eq1}. This trivial mathematical truth will
have great significance when translated to DD physics.  In this
field, $S$ are straight lines and so within \SD , $G(x,y,z,t)$
constitutes a well-defined geometrical wave.  Since energy, by
definition, does not cross streamlines,  \SD constitutes a `flux
tube'~\cite{ref8} carrying the field's total energy without any
loss whatsoever. \SD is a non-diffracting field. It is important
to stress that the border between \SD and \SO does not form a
physical barrier to restrain or reflect the waves, such as the
walls of a waveguide~\cite{ref20} nor is there any discontinuity
in the medium between \SD and \SO such as a change in the index of
refraction: quite simply the amplitude of the waves just drops to
zero in certain regions \SO.

Although mathematically possible, such truncated non-diffracting
fields do not ordinarily exist in nature~\cite{ref9}: the moment
any attempt is made to truncate an infinite field by placing in
its path an opaque screen with an open aperture forming a
cross-section of \SD, the field streamlines $S$ automatically
swerve behind the screen, slow down near the edge and cause
interference effects with regions of zero and maximum intensities,
and transform the whole field topologically in a typical
diffraction pattern. The field now spreads to fill all of space
anew, but now no longer as a geometrical field. Similarly, if the
field is emitted from an infinite number of point sources arrayed
in an aperture \SD, it diffracts as if a plane wave had passed
normally through an open aperture \SD.

\begin{figure}[htb]
\includegraphics[width=8.6cm]{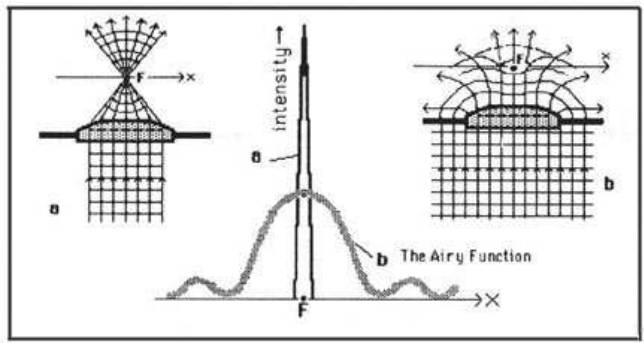}
\caption{\label{fig:2}(a)~Truncated plane wave super-focused at F,
and its spike-like intensity function. (b)~An infinite plane wave
focused  and diffracted  an extended Airy intensity function.}
\end{figure}

Were truncated geometrical fields to exist, they would be very
useful indeed. As shown in Fig.~2(a) if a field is passed through
a focusing system whose every aperture and field stop is larger
than \SD and so does not disturb the edges of the field, it
continues to act as a geometrical field and can be theoretically
focused to a point with infinite intensity and resolution in
imaging systems. Similarly, when the $S$ are parallel, the
conceptual flux tube carries a beam of infinite length and uniform
cross-section \SD with no loss of intensity. The ``diffraction
limits" now used as a measure of quality in focusing instruments
and beams, and which depend on the wavelength, and inversely on
the aperture diameter, would cease to have any significance.

\section{Geometrical Fields Diffract}\label{sec:Sec3}
In hydrodynamics,
\begin{equation}\label{eq4}
\Omega= \text{curl \textbf{S}}
\end{equation}

\noindent{is} known as the vorticity  of the field: it is the
measure of how much the field curves around during its flow. When
$\Omega=0$ as in Eq.~\ref{eq2}, the field is then said to be
irrotational~\cite{ref10} and the vorticity is the same on all of
$S$. When such geometrical fields are interrupted or disturbed by
a physical obstacle, however, the field acquires vorticity around
the edges of the aperture, with $S$ curving into the shadow
regions. This is easy to understand by considering sound waves
`turning' and being heard behind the corner of a building.
Similarly, in hydro- and aerodynamics, obstacles such as the
aperture edge can create wakes, vortices and other deviations from
geometrical flow~\cite{ref11}. In  wave fields where
$\lambda\neq0$ it is more common to describe these changes as
diffraction effects, but in fact I have attempted to show that the
two classes of phenomena are closely linked~\cite{ref5,ref6}. An
extensive description of the electromagnetic field's angular
momentum, using hydrodynamical concepts has been made by
D.~Ximing~\cite{ref12}.

\begin{figure}[htb]
\includegraphics[width=8.6cm]{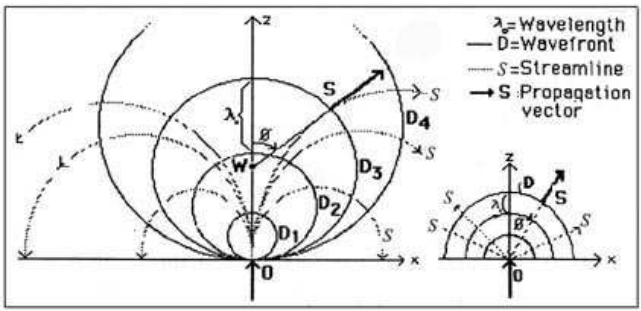}
\caption{\label{fig:3} (Left)~A dipole or `bow-wave' model of an
elementary point field emitted from the origin. (Right)~A
Huygens-Fresnel wavelet.}
\end{figure}

\begin{figure}[htb]
\includegraphics[width=8.6cm]{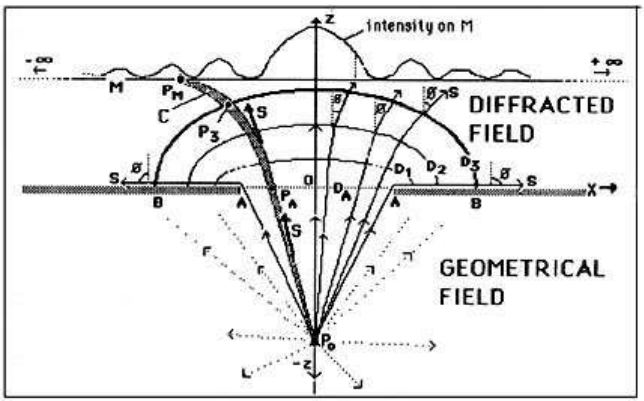}
\caption{\label{fig:4} A geometrical field diffracts.}
\end{figure}

Why do these transformations occur? There is always
\textit{matter} at the edge of a source or an obstacle, where
diffraction starts forming. As shown in Fig.~4, the angle $\phi$
which \bS makes with the normal (the original direction of
propagation rotates by $\pi/2$ or more at the edge. Adjoining \bS
follow suit in a domino effect of decreasing $\phi$ until the
mainstream $S$ at the center of a symmetrical aperture is reached,
where $\phi=0$. This curvature of $S$ around an edge is clearly
illustrated by plots of the streamlines and the ellipsoidal
wavefronts normal  to them, made by Braunbeck and
Laukien\cite{ref13} based on Sommerfeld's~\cite{ref14} rigorous
solution of the infinite half-plane diffraction problem. This
solution also provides another way of visualizing vorticity around
the aperture edge: the concept of a cylindrical wave 'emitted' by
the edge, and interfering with the geometrical field to create the
diffracted field.

This analysis alternatively gives the diffracted field as an
integration of an infinite spectrum of plane waves emitted from
the aperture at angles $\phi$ rotating through $\pi$~\cite{ref15}.
This lends itself naturally to the following physical
interpretation: an infinite number of streamlines normal to the
aperture make up the original zero-order geometrical field
approaching from the $-z$ region. Diffraction bends these $S$ into
an infinite number of streamlines each at a different $\phi$, to
make up the Fourier orders ($-\pi\leq\phi\leq\pi$) of the field in
the $+z$ region~\cite{ref16}.

Kirchhoff's theory of diffraction~\cite{ref17}, although less
rigorous than that of Sommerfeld~\cite{ref14} provides the perfect
theoretical framework for demonstrating the transformation of a
geometrical field into a diffracted one and then, as shown below,
back to a de-diffracted field. A point source $P_0$ on the $-z$
axis in Fig.~4 emits a geometrical spherical wave~\cite{ref18}
whose $S$ coincide with the radii centered on $P_0$. This wave
then passes through an aperture $\overline{AA}$ in an infinite
opaque screen and centered on the $x-y$ plane, and spreads
throughout the $+z$ portion of space, creating the typical maxima
and minima observed on a screen $M$ placed normal to the $z$-axis
at some distance from the aperture. Kirchhoff's method is based on
potential theory and is both inaccurate in the vicinity of the
edges $A$ and noncommittal on the \textit{path} the field's energy
takes from $P_0$ to $M$.  However, the bending of the streamlines
into the shadow regions and the transformation of the geometrical
wavefront $D_A$ into the elliptical wavefronts $D_1,D_2,\ldots$ is
also quite clear in the rigorous solutions mentioned
above~\cite{ref13}.

A typical flux tube $C$ carries energy from $P_0$ to $P_A$ in the
aperture plane along a straight line. But in the $+z$ region it
curves towards the shadow regions behind the obstacle, crossing a
typical wavefront $D_3$ at $P_3$ and creating point $P_M$ in the
diffraction pattern on $M$.  $C$ acts like a channel of a given
capacity~\cite{ref19}, carrying energy from $\overline{P_0P_A}$ to
$P_M$ . This idea is confirmed by Helmholtz' reciprocity (or
reversion) theorem~\cite{ref20} whereby ``a point source at $P_0$
will produce at $P_M$ the same effect as a point source of equal
intensity placed at $P_M$ will produce at $P_0$''.  When $P_0$ is
moved down on the axis to $z = -\infty$ the geometrical field in
the $-z$ regions becomes an infinite plane wave arriving normally
at $\overline{AA}$. The aspect of Kirchhoff's integral that is of
interest here is that the diffracted field reduces to an
integration of Huygens-Fresnel (HF)~\cite{ref21} wavelets emitted
by point sources on $\overline{AA}$. It will be axiomatic to
extend this idea to say that a \textit{ray} from $P_0$ reaches the
aperture plane at a \textit{point} $O$ (a source anywhere on the
aperture) and is transformed into the `exploding' diffracted
pattern of the HF wavelet of Fig.~3 with its inclination factor
adjusting the amplitude by $X=(1+\cos\phi)/2$; but this is known
to be only an approximation.

Since the precise determination of the wavefronts near the
aperture is an essential first step in the implementation of DD
theory, an attempt will now be made to examine the wavelet more
closely.

\section{The Bow Wavelet}\label{sec:Sec4}
A pebble is dropped in a pond and a circular ripple results. But
when an atom emits a photon, or when a point element of a field is
examined, there is radial momentum but no backward momentum such
as in the HF wavelet. Recently Miller~\cite{ref22} starting out
with the wave equation derived a ``spatio-temporal dipole'' field
for the fundamental  electromagnetic wavelet. This confirmed my
independent intuitive derivation~\cite{ref23} adapting the model
of a bow wave such as that made by a boat~\cite{ref24}: a
stationary atom releases a photon having both radial and foreword
velocities $r$ and $f$ and the bow wave or dipole pattern of
Fig.~3 emerges. That some such process is at work is suggested by
the case of Cerenkov radiation, where a fast-moving electron in
glass creates a V-shaped wake of light~\cite{ref25}. In free
space, however, since $r=f=c/2$ the center $W$ of the bow waves'
circular wavefronts $D$, travels up the $+z$ axis at the same rate
as the radius $\overline{WP}$ expands, and the V-shaped
`shock-wave' now opens up to become a straight line along the
$x$-axis. This pattern is also confirmed in the literature, where
the streamlines of a dipole~\cite{ref26} are similar to such
circles centered along the $x$-axis. But on this axis, apart from
the origin, the field is zero since the contributing wavefronts
there would have an infinite radius. The circular streamlines of
the bow wavelet provide a natural explanation for the vorticity of
the field at the edge of the aperture. In addition the bow wave's
wavelength $\lambda(\phi)$ is $\lambda_0$ only in the propagation
direction (unlike the HF wavelet which has only amplitude changes
with $\phi$. $\lambda(\phi)$ explains the very rapid fluctuations
of the diffracted field very near the edge~\cite{ref13}. This
discussion of the primary wavelets of a field is to establish a
method to derive the mathematically exact form of the extended
wavefronts near the aperture, in order to make DD precise. In
section~VI  experiments will be described that show that the bow
wave and the ray that creates it can also have a real physical
existence of their own.

\section{DD: Reversal of the Diffracted Field}\label{sec:Sec5}
It has been long known that electromagnetic fields are time
invariant~\cite{ref27}: any solution $G(x,y,z,+t)$ of Maxwell's
equations will have an equivalent solution $G(x,y,z,-t)$ since
replacing $t$ with $-t$, and hence $v$ with $-v$ will lead to the
same wave equation Eq.~\ref{eq1}. Examples of this are spherical
waves and plane waves whose complete solutions show identical
waves travelling with opposite velocities $\pm v$, for example the
plane wave $G(x,y,z,t)= m(wx+vt)+n(wx-vt)$~\cite{ref28}. Time
invariance here need not involve any philosophical considerations
of the arrow of time, merely that the field retains all its point
values when the velocity direction is reversed by rotating all its
\bS by $\pi$. The principle of reversibility along rays has also
been derived from Fermat's principle~\cite{ref29}.

In optics this principle of reversibility is implicit in any
imaging situation where a point object and its image can be
interchanged, and in the theory of holography~\cite{ref30}. With
this in mind consider two typical situations where diffraction
occurs: First the isolated bow wavelet of Fig.~3.  A straight ray
is scattered at a point $O$ and  then branches into streamlines,
creating any of the diffracted wave fronts $D_1, D_2, D_3,\ldots$
Now imagine that a curved mirror shaped as one of the wavefronts
say $D_2$, is placed to overlap that wavefront exactly. Since all
the diffracted streamlines are then normal to the mirror the \bS
vectors are reflected back on themselves, and since amplitude and
phase are preserved (phase changes upon reflection will be uniform
across $D$) then the field will reverse itself along the same
streamlines, and passing point $O$ on the return trip, ends up as
a single ray in the $-z$ direction (but without the original
source of the ray there) . Alternately, instead of a mirror think
that $D_2$ now consists of an infinite array of point sources with
an identical amplitude function to that of the original diffracted
pattern on $D_2$. In the $+z$ direction the field will continue to
diffract exactly as before. But below $D_2$ in the $-z$ direction
the sources on $D_2$ will recombine to create a reversing ray. The
second case is that of an extended field passing through an
aperture, for example the diffraction situation of Fig.~5.

\begin{figure}[htb]
\includegraphics[width=6cm]{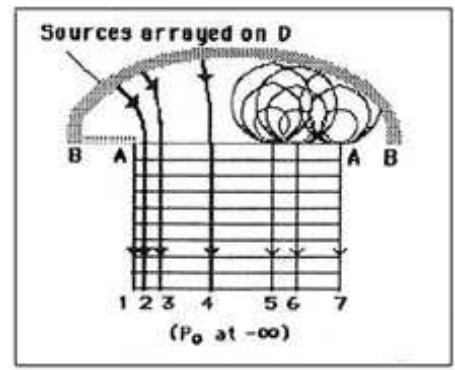}
\caption{\label{fig:5} De-diffraction (DD) by reversal of a
diffracted wavefront.}
\end{figure}

Here again any one of the wavefronts $D$ of Fig.~4 can serve as a
source for the reverse DD field, or to reflect an already
diffracted field. As a result of these procedures, the curved
streamlines $1,2,3,\ldots$ of the reversing field emerge from the
aperture $\overline{AA}$ normally as a part of a dediffracted
truncated geometrical field. If diffraction is to be likened to a
topological distortion of an ordinary field, the reverse process
can be said to start with a purposely distorted field to produce
an ordinary one. It might be argued that the presence of an opaque
screen $\overline{BA},\overline{AB}$ reflecting or absorbing
portions of the original $-z$ field will necessitate the placement
of a similar screen when DD is attempted. Such a screen is not
needed for DD however, because the diffracted field immediately
touching the screen from the $+z$ side will be negligible or zero;
indeed that the field is zero on the screen is one of the boundary
conditions in Kirchhoff's derivation~\cite{ref17}. Moreover, when
the original source is an infinite array of point sources along
$\overline{AA}$, no screen $\overline{BA},\overline{AB}$ need be
present in the diffracting field, and none will be needed in the
reversing field, only the conceptual removal of the original
sources on $\overline{AA}$ to allow the DD field to emerge.

Choosing the wavefronts as the source for a DD field insures that
the phase is the same there, simplifying the design of the lenses
or antennas to be used to create a DD field. However in principle
\textit{any} linear (or surface in the 3D case) array $M$ of
sources spanning the diffracted field and mimicking its local
phase and amplitude can be used to create the DD field. For
example an infinite array of sources having the ring-shaped phase
and amplitude pattern of the electric field of the Airy
function~\cite{ref31}, and placed in the focal plane of a lens
will cause a reverse DD field to emerge from the lens. This method
greatly resembles the zone-plate-type filter described by Toraldo
di Francia to achieve superresolution~\cite{ref2,ref3}. Of course,
in any practical application, such a filter will have a limited
size, and the imitation of the whole diffracted field, and hence
of  DD, will fail. A more practical filtering DD method would be
to wrap a rigid curved holographic film to roughly encircle an
aperture, and illuminate both its $z$ sides, in the presence of
the aperture, with coherent plane waves. The developed illuminated
plate will then recreate the DD field when illuminated from the
$+z$ side.

In practice a DD field can be accomplished by any one of five
general procedures: 1)~An array of sources mimicking the local
phase and amplitude values along an entire continuous  random
cross-section $M$ of  the original field. 2)~By reflecting a
diffracted field normally upon itself. 3)~The creation of the
wavefronts $D$ by suitably focusing a plane or spherical wave.
4)~Holographic methods or 5)~A filter illuminated normally by a
plane wave (hence the phase is uniform on the filter) having only
amplitude changes. Finally, it must be cautioned that when
performing DD field calculations, using elementary wavelets, or
the Kirchhoff integral, these must be used in the correct
orientation. The wavelets always 'point' from the geometrical rays
5,6,7 of Fig.~5 to the circular diffracted patterns. An
integration of bow wavelets on $\overline{AA}$ will give $D$, but
integrating wavelets on $D$ will not give the field on
$\overline{AA}$ or the DD field in $-z$, since on $D$ the field is
\textit{already} diffracted. A rule of thumb is that the Kirchhoff
integral should not be used to evaluate the field of wavefronts
such as $D$, whose normals \bS rotate rapidly through $\pi/2$ at
the aperture edge.

\section{De-Diffraction Experiments}\label{sec:Sec6}
Simple experiments to prove the DD methods outlined in this paper
were performed as follows, but a more sophisticated experimental
verification of DD using electromagnetic radiation is still
needed.

\begin{figure}[htb]
\includegraphics[width=8.6cm]{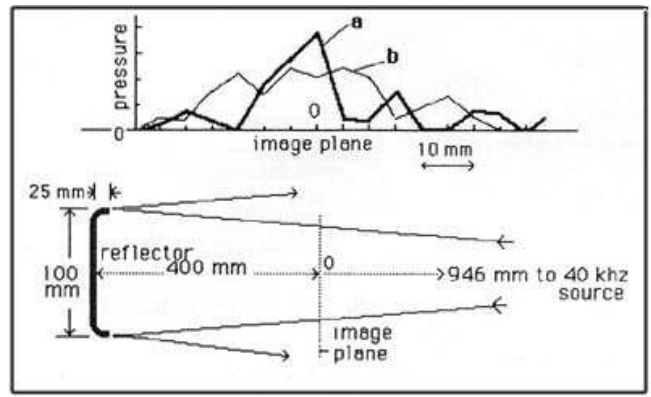}
\caption{\label{fig:6}Ultrasound wave field measured after
reflection from (a)~DD plane reflector, and (b)~From a plane
reflector of the same size.}
\end{figure}
DD OF ULTRASOUND WAVES:  The field of a 40 kHz ultrasound source
as reflected from a plane antenna and one with curved edges were
compared as in Fig.~6. The curved edges, roughly approximating the
curvature of a DD wavefront, concentrated the field appreciably.

\begin{figure}[htb]
\includegraphics[width=8.6cm]{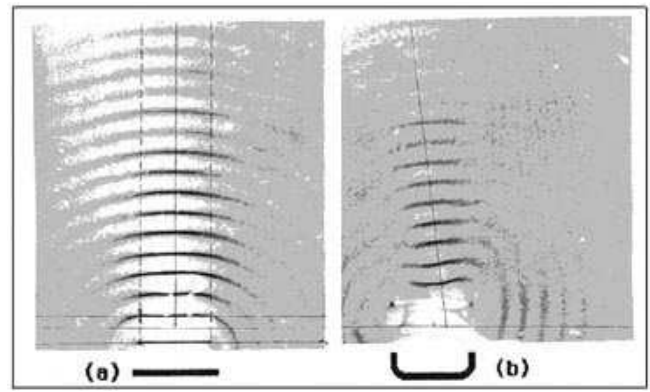}
\caption{\label{fig:7} Photographs of waves emitted in a ripple
tank from (a)~A vibrating plate reflector three wavelengths wide,
and (b)~From its DD equivalent.}
\end{figure}
WATER RIPPLES:  Water ripple experiments are shown in Fig.~7: a)~a
vibrating flat plate produced ripples of a wavelength one third
its width spreading out in the typical oval diffracted wavefronts
(b) roughly curving the edges through a quarter-circular arc of
about one $\lambda$ radius produced a dramatic concentration of
the waves mainly in the forward direction. Note the absence of
waves from either side of the axis in (b): the waves in the bottom
right of  (b) are spurious, since they are emitted from the back
of the plate.

\begin{figure}[htb]
\includegraphics[width=8.6cm]{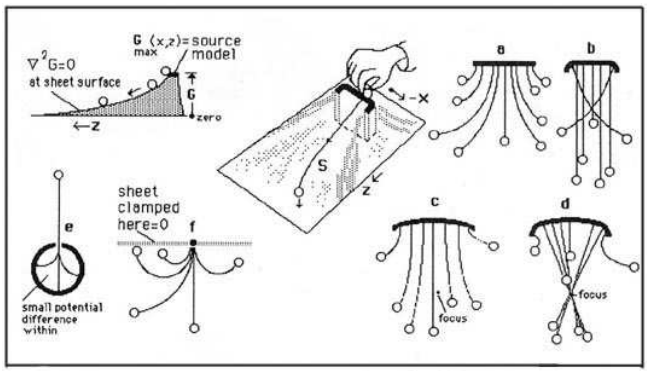}
\caption{\label{fig:8} Diagrams of gravity potential field
experiments.  The path taken by a marble rolling down a flexible
sheet stretched over elevated forms,  simulates the streamlines S
of (a)~A diffracted plane wave, (b)~DD plane wave, (c)~Diffracted
focused wave, (d)~Focused wave, (e)~DD (reversed) bow wave and
(f)~Bow wave. }
\end{figure}
GRAVITATIONAL MODEL. It is known that a thin flexible rubber sheet
stretched over a solid two dimensional horizontal model of a
certain function $G(x,y)$ satisfies Laplace's equation of $G$:
Freezing a wave in time reduces the right hand term of
Eq.~\ref{eq1} to zero, and the wave equation becomes $\nabla^2
G=0$. The surface of the sheet represents both the gravitational
and electrostatic potentials of this Laplacian~\cite{ref32}. For
our purposes, the potential of the fields we are studying was
represented by a flat model $G(x,z)$ of the wavefront in question.
The model was lifted a distance $G_{max}$ representing the initial
potential of the wave and the sheet stretched over the model as in
Fig.~8.

The gravity-wave-potential equivalence mentioned above (and which
might not be coincidental, as discussed below) allows a study of
the streamlines $S$ of the field since they will be the path taken
by a marble rolling down the sheet's surface. The equipotential
horizontal contours of the sheet's surface ($G_1,G_2,G_3,\ldots$
the `elevation') are then the wavefronts. $G_{max}$ is
proportional to the energy of the source (inversely proportional
to the wavelength) and it is interesting to see how increasing the
height of the model reduces the diffraction spreading, according
to theory. This method though not too quantitatively exact if the
membrane slope is larger than around $18^\circ$, gives an
excellent idea of the physical situation. As shown in the
streamline sketches of Fig.~8, such a gravitational model was used
to prove that DD proceeds exactly as theoretically predicted for a
diffracted and de-diffracted plane (a,b) and focused (c,d) waves.
The concentration of the streamlines at a single point-focus in
case (d) was a dramatic demonstration of superresolution. The bow
wavelet pattern both in the emission (f) and the DD reversing (a)
modes (a lone ray) was also verified. In the case of the point
source (f) the sheet was also clamped down ($G=0$) along $x$
except for a spike at the origin, because the source was assumed
to be located on a flat non-conducting plane. The paths of the
rolling marble in (f) were very similar to the streamlines of the
bow-wavelet $S$ of Fig.~3.

\section{Practical Applications of DD}\label{sec:Sec7}
The cancellation of diffraction improves the performance of a wide
range of instruments in many fields. These can be roughly
categorized according to whether the truncated field is to be
focused or allowed to propagate as a beam. And also according to
the type of field such as optical, microwave, acoustic etc.
Examples are too numerous to list, but focused field applications
include cameras, telescopes (optical and radio) microscopes
(optical and electron), imaging radar, and sonar. Beam
applications include lasers and tele-communication parabolic
antennas. In the case of imaging radar, for example, DD
methods~\cite{ref5,ref6}, applied through refocusing the field by
reshaping the antenna by curving its rim, will allow fine
resolution even with a large wavelength and a small antenna, since
a truncated geometrical field can be focused to a point regardless
of the field's wavelength or the aperture size. An optical laser
passing through a DD lens with curved edges pointed towards the
moon should proceed with no divergence (apart from atmospheric
degradation), reaching its target with its original profile
relatively intact. Normally diffraction would spread a 10 cm.
diameter laser beam to some two kilometers in diameter by the time
it arrives there.

\section{Diffraction, Quantum and relativistic Fields}\label{sec:Sec8}
Diffraction has been cited by Heisenberg himself (33) as an
example of uncertainty relations. The product of a photon's
momentum and position cannot be less than Planck's constant $h$.
The question now arises: are uncertainty and hence quantum effects
cancelled together with the cancellation of diffraction? The
inescapable conclusion is that they are, since both momentum and
position would be uniquely known in a DD geometrical field. The
vectors \bS of a plane DD field all point in the same direction,
and the position is always known within \SD  especially in the
case of a single ray. In the bow wavelet itself the change from a
ray (position and momentum precisely known) to the spread-out
diffracted wavelet, with its momentum vectors fanning out at all
angles is a model of this transformation from a classical field
(the ray) to a quantum field (the fan) and vice versa. Are
elementary particles emitting bow wavelets the physical
machinery~\cite{ref34} behind quantum effects? Two particles $P_1$
and $P_2$ emit bow waves which meet at a random angle
$(\phi_1-\phi_2)$ and the resulting interference pattern is taken
for that of quantum probability functions. Here then is the
physical `cause' of quantum effects: it is the variable energy
content transmitted by streamlines or flux tubes from $P_1$ and
$P_2$ to a nearby point $T$, which can be considered as the
probability amplitudes~\cite{ref35} $<P_1|T>$ and $<T|P_2>$.

\begin{figure}[htb]
\includegraphics[width=8.6cm]{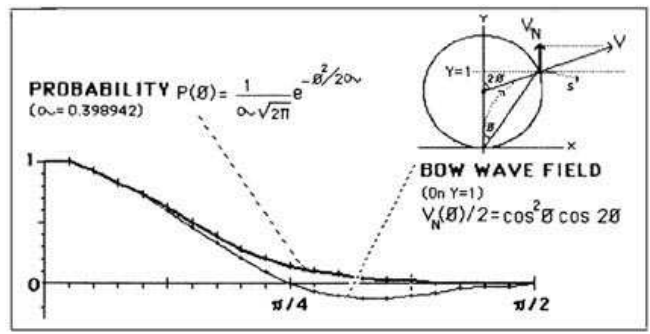}
\caption{\label{fig:9} Bow wave and probability functions. }
\end{figure}

\begin{figure}[htb]
\includegraphics[width=5cm]{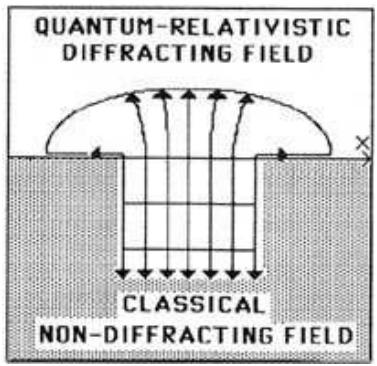}
\caption{\label{fig:10} The transformation of a classical field
through diffraction into a field with quantum and
general-relativistic characteristics.}
\end{figure}
Fig.~9 shows the similarity between a Gaussian probability
function~\cite{ref36} and the normal of the electric field of a
bow wavelet around the propagation axis. Discussions following
Toraldo di Francia's paper on super-resolving filters~\cite{ref2}
also raised the question of quantum uncertainty in diffracted
fields, and included this remark: ``the only correct quantum
electrodynamical version of Heisenberg's principle imposes no
relevant restrictions on resolving power to begin with.'' One can
speculate even further. The emerging circular streamlines of the
bow wavelet first propagate foreword and then `fall' towards the
source like a fountain, as if attracted towards the origin. Can
the streamlines and equipotential surfaces of the primary wavelet
then be interpreted as curved Gaussian coordinates (\bS being the
geodesics~\cite{ref37}) of a relativistic gravitational distortion
of the field surrounding the atom emitting the photons? This would
be a miniature version of Einstein's  results concerning the
bending of light in the vicinity of a massive star~\cite{ref38}.
Considering the proportionally smaller atomic masses and distances
involved, this might not be too improbable. The distortion of
space in the vicinity of an obstacle can be easily observed by
moving a pinhole or slit with an aperture of about 1~mm in front
of the eye. The aperture seems to act like a concave lens
distorting the view, according to the divergence of the diffracted
streamlines~\cite{ref39}. Thomas Young had also combined
gravitational and optical concepts to explain diffraction. He
hypothesized that the rays refract through a lens-like,
increasingly dense ethereal fluid surrounding the obstacle,
``attracted to particles of gross matter"~\cite{ref40}. In view of
the preceding analysis, the intriguing possibility exists that
sub-atomic bow waves propagated into space can be considered the
source of a unified field combining quantum,
relativistic-gravitational and electromagnetic effects. In that
case field reversal methods such as DD can provide a way to
transform quantum-relativistic fields in space into classical
ones, and vice versa, as in Fig.~10. But all such speculation must
first await acceptance and further experimental proof of DD.

\section{Summary and Conclusion}\label{sec:Sec9}
A method was presented, starting from the wave equation, and a
truncated version of a geometrical field, and the principle of
reversal of wave fields, whereby de-diffracted geometrical wave
fields can be created. To clarify the process, the conversion of a
ray into the diffracted primary wavelet and back to a ray was
studied as a model of such a DD transformation. DD methods allow
superresolution in imaging instruments and for infinite beams with
no divergence. The simple experiments performed to prove DD could
be repeated and refined. Rigorous computer simulation of the field
can find the precise waves D for a  given device and thus provide
the exact design of a DD lens, antenna or other instrument. The
similarity of the field equations of flow fields, gravity and
electrostatics suggest that methods equivalent to DD might also be
used to modify those fields too. For example a boomerang-shaped
object would have its gravitational field lines 'focused' on the
concave side. Finally it was speculated whether diffraction could
be just one manifestation of a united quantum-relativistic field
surrounding an atom, and conversely whether DD means  that it is
possible to create classical-field regions of  space free from
such effects.

\begin{acknowledgments}
Thanks are due to Prof.~C.~Shepherd, Dr.~D.A.B.~Miller, and
Mr.~V.~Alexander for correspondence clarifying some questions. And
to Capt.~N.~Kobori for building the ultrasound equipment used for
the DD experiments and to Mrs.~K.~Tamari for help in the design
and performance of the gravity field experiments.
\end{acknowledgments}



\begin{thebibliography}{99}
\bibitem{ref9}
But occasionally, a truncated ripple pattern, similar to the
experimental one in Figure 7(b) has been observed in a stream,
'focused' by a chance arrangement of pebbles and leaves. Also a
primitive attempt to create DD sound waves is when a person
instinctively cups his or her hands around the mouth to
concentrate the emerging waves when shouting to a distant hearer.
\end{thebibliography}
\end{document}